\documentclass[aip,amsmath,amssymb,reprint]{revtex4-1}

\usepackage{graphicx}
\usepackage{dcolumn}
\usepackage{bm}

\usepackage[utf8]{inputenc}
\usepackage[OT1]{fontenc}
\usepackage{tipa}
\usepackage{mathptmx}
\usepackage{etoolbox}

\usepackage[euler]{textgreek}
\usepackage{siunitx}

\draft 

\begin{document}




\title[Ring QCSEL]{Microring quantum cascade surface emitting lasers}

\author{David Stark}
\email{starkd@phys.ethz.ch}
\affiliation{Institute for Quantum Electronics, Department of Physics, ETH Z\"urich, 8093 Z\"urich, Switzerland}





\author{Mattias Beck}
\affiliation{Institute for Quantum Electronics, Department of Physics, ETH Z\"urich, 8093 Z\"urich, Switzerland}

\author{J\'{e}r\^{o}me Faist}
\email{jfaist@ethz.ch}
\affiliation{Institute for Quantum Electronics, Department of Physics, ETH Z\"urich, 8093 Z\"urich, Switzerland}

\date{\today}


\begin{abstract}
    We miniaturize a vertically-coupled in-plane whispering gallery mode cavity incorporating a quantum cascade gain medium, aiming to realize the mid-infrared counterpart to the vertical cavity surface emitting laser (VCSEL). 
    Building on previous work with linear microcavities, we introduce a new type of quantum cascade surface emitting laser (QCSEL) by miniaturizing a buried heterostructure ring cavity.
    At wavelengths of \qty{4.5}{\um} and \qty{8}{\um}, we investigate the optical losses for decreasing ring diameters while benchmarking the device performance against linear microcavities.
    We achieve an equivalent mirror reflectivity of 0.95 and demonstrate lasing with ring diameters as small as \qty{50}{\um}.
    Finally, we report a continuous-wave threshold power dissipation of \qty{274}{mW} for a \qty{100}{\um} diameter ring QCSEL, characterized on wafer level at \qty{20}{\degreeCelsius}.

\end{abstract}


\maketitle 

\section*{Introduction}
    Pursuing a low threshold single-frequency laser on a compact footprint, cavities that support whispering gallery modes (WGMs) are excellent candidates.\cite{McCall1992WhisperinggalleryLasers}
    In a WGM cavity, light is laterally confined by a closed loop concave boundary, and it is guided along this boundary through total internal reflection.\cite{He2013WhisperingLasers}
    Consequently, the light is effectively trapped inside the cavity and it couples only evanescently to its surrounding medium. Additional losses may occur due to surface roughness at the boundary resulting from fabrication imperfections which scatter the circulating light.
    In contrast to linear cavities, where high reflectivity (HR) mirrors are technically challenging to fabricate, a WGM cavity benefits from the concave boundary inherently providing a mirror with virtually unity reflectivity, leading to low optical losses.
    Depending on the refractive index contrast realized at this boundary extremely tight bending radii are feasible, enabling miniature cavity volumes supporting low-power operation.
    Additionally, as the free spectral range of the cavity modes scale inversely with the cavity perimeter, miniaturization offers a natural way to achieve single-mode selection.
            
    Since the early days of the QCL, numerous studies have implemented MIR QCLs in WGM cavities.\cite{Kacmoli2024QuantumLasers}
    Initial studies demonstrated single mode operation by reducing the diameter of microdisk lasers at cryogenic temperatures.\cite{Faist1998QuantumLasers,Gmachl1997Long-wavelengthLasers} 
    Microdisks with diameters between \qty{17}{\um} and \qty{125}{\um} were studied and the lasers emitted at wavelengths near \qty{5}{\um}\cite{Faist1998QuantumLasers} as well as around \qty{9.5}{\um} and \qty{11.5}{\um}.\cite{Gmachl1997Long-wavelengthLasers}
         
    Another type of WGM cavity, which has been extensively studied in combination with QCLs, is the grating-coupled surface emitting ring QCL with diameters on the order of \qty{400}{\um}.\cite{Mujagic2008Grating-coupledLasers, Mujagic2008LowLasers}
    The ring-shaped ridge QCL incorporates a resonant second order grating that is dry-etched into the top InP cladding.
    This surface grating, formed by the refractive index contrast between air and InP, enables the vertical outcoupling and provides optical feedback to select a single mode.
    With such a device architecture and about \qty{10}{\um} wide ridges, single frequency lasing has been demonstrated up to \qty{380}{K} in pulsed operation and up to \qty{230}{K} in continuous wave (CW) operation.\cite{Mujagic2010RingLasers}
    The versatility of this device architecture has been highlighted by a two-dimensional array of ring QCLs.\cite{Mujagic2011Two-dimensionalArrays}
    By utilizing distinct grating designs, the array emitted 16 discrete wavelengths across a \qty{180}{cm^{-1}} bandwidth centered at \qty{8.2}{\um}.
    Subsequently, the substrate emission for different grating period duty-cycles have been investigated.\cite{Schwarzer2012GratingLasers}
    By introducing $\pi$ phase shifts in the grating, a central lobed far-field was realized. \cite{Schwarzer2013LinearlyLasers}
    Furthermore, a large variety of different grating or substrate arrangements modifying the far-field properties have been reported.\cite{Szedlak2014Grating-basedLasers,Szedlak2014On-chipLasers,Szedlak2015TheLasers,Szedlak2016RingSpectroscopy}
    However, for these large-area and ridge-type QCLs, heat extraction remains challenging, hampering CW operation at elevated temperatures.
    \begin{figure*}[t!]
        \centering
        \includegraphics{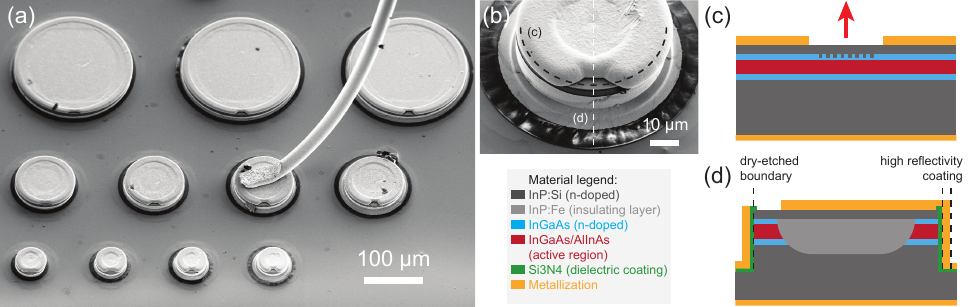}
        \caption{
        The ring QCSEL device architecture: 
        (a) Scanning electron microscopy (SEM) image displaying of three rows of devices after fabrication. The devices in the top, middle, and bottom rows feature diameters of \qty{200}{\um}, \qty{100}{\um}, and \qty{50}{\um}, respectively.
        (b) SEM image providing a magnified view on the metallic aperture of a device with \qty{50}{\um} in diameter. The dashed lines display the cross-section in both the azimuthal and radial direction.
        (c) Schematic diagram illustrating the cross-section in the azimuthal direction along the arcuate dashed line depicted in (b).
        (d) Schematic diagram illustrating the cross-section in the radial direction along the vertical dashed line shown in (b).
        }
        \label{fig:ring_concept}
    \end{figure*}
    
    Room-temperature CW operation has been successfully demonstrated with substrate-emitting buried heterostructure ring QCLs using diameters of \qty{800}{\um}.\cite{Bai2011HighLaser,Wu2017HighOperation}
    For the initial demonstration a buried semiconductor grating, where the index contrast was formed by InGaAs and InP, in the vicinity of the waveguide was used.\cite{Bai2011HighLaser}
    In the subsequent demonstration, the grating was formed by Ti/Au and InP.\cite{Wu2017HighOperation} For both demonstrations, the second order grating was exploited for vertical outcoupling and in-plane optical feedback. Additionally, the devices were soldered with the epitaxial layer down onto AlN submounts, allowing an effective heat extraction.
    A drawback of these buried heterostructure WGM cavities is that the outer boundary provides a relatively low refractive index contrast $(\sim 0.2)$ and therefore these cavities are limited to large bending radii ($\gtrsim$ \qty{100}{\um}).          
            
    Recently, edge-emitting elliptic disk WGM microcavities have been investigated at emission wavelengths near \qty{8}{\um}\cite{Guo_8um} and near \qty{4.7}{\um}.\cite{Guo_4p5um}
    Following the original demonstration,\cite{Wang2010Whispering-galleryAction} these WGM microcavities employed an elliptic boundary with a minor radius of \qty{80}{\um}, a major radius of \qty{96}{\um}, and a wavelength-sized notch in the elliptic boundary to support directional in-plane lasing emission.
    Guo and co-workers\cite{Guo_8um, Guo_4p5um} advanced the original WGM microcavity.\cite{Wang2010Whispering-galleryAction}
    In particular, they incorporated a selective insulating InP top cladding to reduce the electrical dissipation and applied a SiO\textsubscript{2} coating to passivate the surface on the outer boundary.
    While the CW operation at longer wavelengths was limited up to \qty{5}{\degreeCelsius},\cite{Guo_8um} CW operation at shorter wavelengths was achieved up to \qty{50}{\degreeCelsius} relying on epitaxial layer down mounting.

    Here, to minimize the electrical power dissipation and to exploit narrow waveguides allowing efficient heat extraction, we implement a buried heterostructure ring cavity. To achieve small bending radii and avoid any lateral light emission, the outer boundary of the cavity is dry-etched and HR coated on wafer-level.
    Light is extracted vertically from short arcuate grating section near the waveguide core, using 5 or 7 grating periods.
    This device builds upon our work with linear microcavities \cite{Stark2024QuantumLasers} and represents another type of quantum cascade surface emitting laser (QCSEL).
    In the following sections, we introduce further details of the ring QCSEL and examine the reflectivity of the outer boundary.
    We then present a cavity perimeter scaling experiment in pulsed operation, followed by CW measurements, both performed at \qty{20}{\degreeCelsius}.

\section*{Device architecture and reflectivity of the outer boundary}
    For this study, the same fabrication process was followed and the same active regions, EV1464 and EV2616, were used, as described for the linear QCSELs.\cite{Stark2024QuantumLasers} For CW measurements, the active region EV1299, reported by Bismuto and co-workers\cite{Bismuto2011InfluenceLasers,Bismuto2011Mid-infraredLasers}, was additionally used.
    The ring QCSEL device architecture is illustrated in Fig.~\ref{fig:ring_concept}, with the diameter of the outer dry-etched boundary $D$ ranging from \qty{600}{\um} down to \qty{50}{\um}.
    
    \begin{figure}[b!]
        \centering
        \includegraphics[]{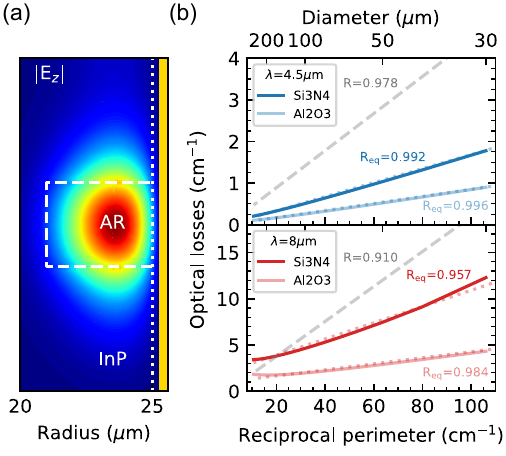}
        \caption{
            2D axissymmetric simulations of radial waveguide cross-sections: 
            (a) The electric field profile $|E_z(r,z)|$ of the fundamental WGM mode of a ring cavity with a dry-etched boundary diameter of \qty{50}{\um}. 
            The active region core (dashed lines) and the InP cladding are annotated. The vertical dotted line indicates the boundary between the dry-etched mesa and the dielectric coating. The yellow stripe illustrates the metallic coating.
            (b) Optical losses deduced from the fundamental WGM mode for varying reciprocal perimeters and different HR coatings. The top and the bottom panel show the simulations performed at a wavelength of \qty{4.5}{\um} and \qty{8}{\um}, respectively. 
            The corresponding diameters of the devices are indicated with the top axes in each panel.
            The dotted lines depict the linear fitting of the optical losses to Eq.~(\ref{eq:mirror}). The resulting equivalent reflectivities $R_\text{eq}$ are annotated.
            The gray dashed lines display the mirror losses of linear cavities using the linear perimeter $p=2L$ and the simulated reflectivities of 0.910 and 0.987.\cite{Stark2024QuantumLasers}
        }
        \label{fig:ring_sim_losses}
    \end{figure}
    \begin{figure*}[t!]
        \centering
        \includegraphics[scale=1]{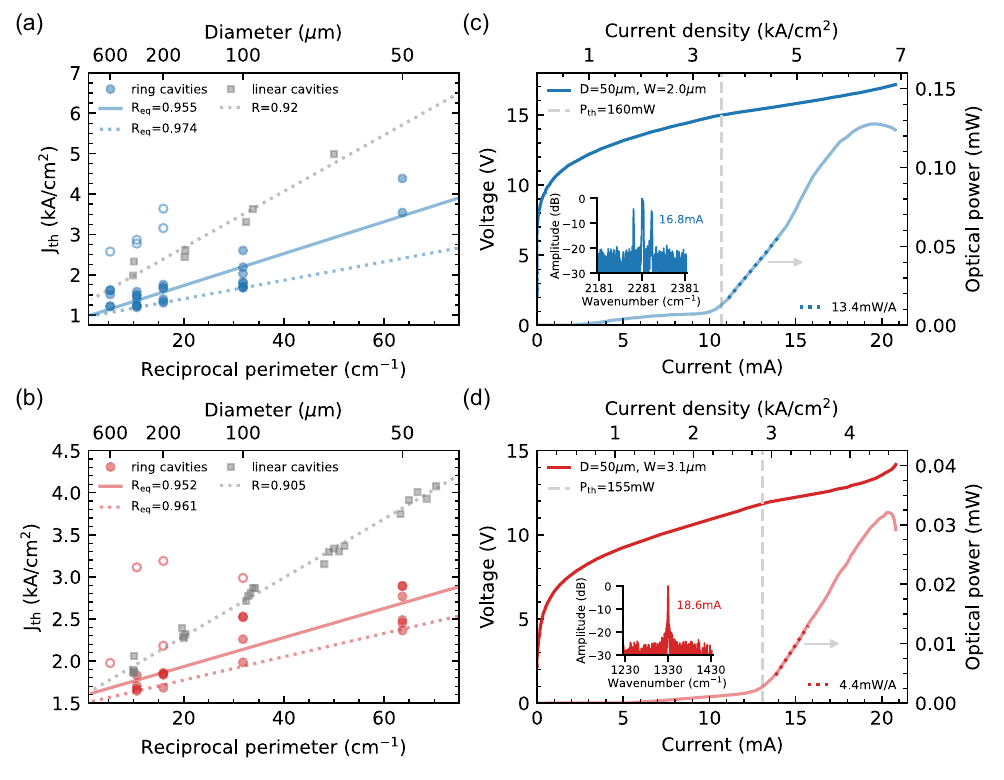}
        \caption{
        Cavity perimeter scaling based on the active regions EV1464 (a) and EV2616 (b):
        The threshold current densities of ring cavities (circles) and linear cavities (squares) are displayed. 
        The top axes indicate the corresponding diameter of the ring cavities.
        The linear curves represent the fitting of $J_\text{th}$ to Eq.~(\ref{eq:jth_ring}).
        The resulting reflectivities are annotated.
        For the fits illustrated with solid lines, all ring QCSELs except the ones indicated with empty circles are included.
        The colored dotted lines depict the fits for best case reflectivities, considering only the lowest thresholds between \qty{300}{\um} and \qty{100}{\um} for EV1464 and between \qty{300}{\um} and \qty{50}{\um} for EV2616, respectively.
        The pulsed LIV characteristics featuring the lowest power dissipation at threshold $P_\text{th}$ of EV1464 (c) and EV2616 (d). The ring diameter $D$, the waveguide width $W$, the slope efficiencies $\eta$, and the dissipation power at threshold $P_\text{th}$ are annotated. The insets show a spectra acquired with a resolution of \qty{0.5}{cm^{-1}}. All the measurements shown in (a) - (d) were performed in pulsed operation at \qty{20}{\degreeCelsius} with a repetition rate of \qty{96.15}{kHz} and a pulse width of \qty{312}{ns}.
        }
        \label{fig:ring_jth_vs_perimeter}
    \end{figure*}
    Targeting miniaturized ring cavities, we investigate the optical losses induced by the outer boundary for decreasing ring diameters.
    These losses are derived from 2D axissymmetric simulations, which compute the fundamental WGM mode of the radial waveguide cross-section, as depicted in Fig.~\ref{fig:ring_sim_losses}(a).
    In Fig.~\ref{fig:ring_sim_losses}(b), the optical losses are plotted as a function of the reciprocal cavity perimeter, expressed as $1/p=1/(\pi D)$. 
    While an exponential increase in optical losses is expected for decreasing perimeters $p$ in devices with purely dielectric lateral confinement, in our case, the losses are dominated by the absorption in the dielectric-metal coating and are well approximated by the mirror losses of an equivalent linear cavity, given by
    \begin{equation} \label{eq:mirror}
        \alpha_\text{m} = \frac{\ln(1/R_\text{eq}^2)}{p},
    \end{equation}
    where $R_\text{eq}$ represents the equivalent reflectivity of both end-mirrors. By fitting the optical losses of ring cavities to Eq.~(\ref{eq:mirror}), the equivalent reflectivities are computed.
    In the case of the fabricated Si\textsubscript{3}N\textsubscript{4} coating, equivalent reflectivities of 0.957 and 0.992 are deduced at wavelengths of \qty{4.5}{\um} and \qty{8}{\um}, respectively.
    Hence, theoretically superior reflectivities can be achieved in comparison to linear cavities, where reflectivities of 0.978 at \qty{4.5}{\um} and 0.910 at \qty{8}{\um} are computed.\cite{Stark2024QuantumLasers}
    Ultimately, even larger equivalent reflectivities are predicted for ring cavities if Al\textsubscript{2}O\textsubscript{3} as the dielectric coating is employed instead of Si\textsubscript{3}N\textsubscript{4}, as shown in Fig.~\ref{fig:ring_sim_losses}(b).
    

\section*{Results}
    To evaluate the performance of the ring QCSELs, we measure their threshold for different diameters in pulsed operation.
    The resulting threshold current densities are then compared to those of linear QCSELs, see Fig.~\ref{fig:ring_jth_vs_perimeter}. 
    Lower threshold current densities are observed with ring cavities for both active regions, EV1464 and EV2616, except for a few devices where we suspect fabrication imperfections.
    Given that theoretically lower optical losses are predicted for the ring cavities, we neglect these exceptional devices in the following discussion.
    The threshold current densities of the ring cavities can be expressed as
    \begin{equation}  \label{eq:jth_ring}
        J_\text{th} = \frac{1}{g\Gamma}\left(\alpha_\text{i} + \frac{\ln(1/R_\text{eq}^2)}{p}\right),
    \end{equation}
    where $g$ is the material gain, $\Gamma$ the optical confinement factor, and $\alpha_\text{i}$ the internal losses. Using the material gain values reported previously,\cite{Stark2024QuantumLasers} we fit the threshold current densities to Eq.~(\ref{eq:jth_ring}) to eventually assess the equivalent reflectivities.

    For the active region EV1464, designed for an emission wavelength of \qty{4.5}{\um}, we compute an equivalent reflectivity of 0.955. This corresponds to loss improvement of more than 44 \% with respect to the linear cavities.
    In the best case, considering only the lowest threshold current densities of rings with diameters between \qty{300}{\um} and \qty{100}{\um}, an even higher equivalent reflectivity of 0.974 is suggested. However, both experimentally deduced reflectivities deviate from the simulated value (see Fig.~\ref{fig:ring_sim_losses}), which we attribute to the surface roughness at the dry-etched boundary resulting from fabrication imperfections. 

    For the active region EV2616, designed for an emission wavelength of \qty{8}{\um}, an equivalent reflectivity of 0.952 is deduced.
    Consequently, the losses are improved by more than 50 \% compared to the linear cavities.
    Moreover, when considering only the lowest threshold current densities of rings with diameters between \qty{300}{\um} and \qty{50}{\um}, an equivalent reflectivity of 0.961 is extracted.
    These values agree well with the simulated value of 0.957. 


    \begin{figure}[t!]
        \centering
        \includegraphics{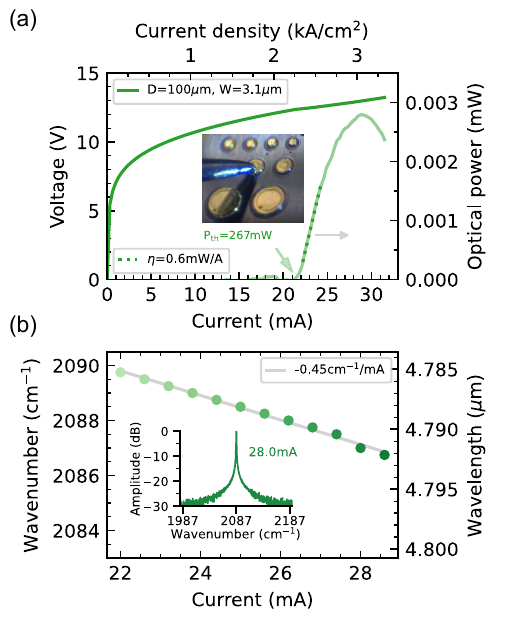}
        \caption{
        Continuous wave operation of a ring QCSEL based on EV1299 characterized on wafer level at \qty{20}{\degreeCelsius}:
        (a) LIV characteristics including annotations for diameter $D$, waveguide width $W$, slope efficiency $\eta$, and threshold dissipation power $P_\text{th}$. The inset shows the wafer-level measurement of a \qty{100}{\um} diameter device.
        (b) Peak wavenumber of the dominant mode throughout the dynamic range of the laser.  The inset shows the spectrum acquired at \qty{28}{mA} and a resolution of \qty{0.5}{cm^{-1}}.
        }
        \label{fig:ring_CW}
    \end{figure}
    With the ring QCSELs, narrow waveguides can be realized in small-diameter rings suited for low electrical power operation.
    Among the characterized devices, the lowest power dissipation at threshold is obtained from \qty{50}{\um} diameter rings with waveguide widths narrower than \qty{3.2}{\um}. 
    The corresponding light-current-voltage (LIV) characteristics and spectra, acquired in pulsed operation, are shown in Fig.~\ref{fig:ring_jth_vs_perimeter}(c) and (d).
    
    For the device based on EV1464 (see Fig.~\ref{fig:ring_jth_vs_perimeter}(c)), the power dissipation is \qty{160}{mW} at threshold and \qty{332}{mW} at maximum optical power. At \qty{16.8}{mA} three modes near a wavenumber of \qty{2280}{cm^{-1}} are observed with a free spectral range of about \qty{20}{cm^{-1}}.
    For the device based on EV2616 (see Fig.~\ref{fig:ring_jth_vs_perimeter}(d)), the power dissipation is \qty{155}{mW} at threshold and \qty{280}{mW} at maximum optical power. The laser emits a single-mode near a wavenumber of \qty{1330}{cm^{-1}} across the entire dynamic range. At a current of \qty{18.6}{\mA} a side mode suppression ratio (SMSR) exceeding \qty{20}{db} is observed.
    From the reflectivity analysis presented in Fig.~\ref{fig:ring_jth_vs_perimeter}(a) and (b), we expect that the overall dissipation for both active regions can further be reduced by fabricating rings with diameters smaller than \qty{50}{\um}.
    The smallest diameter for which CW lasing is observed at \qty{20}{\degreeCelsius} is \qty{100}{\um} for both active regions.
    For EV1464, a threshold power dissipation of \qty{352}{mW} is observed, while for EV2616, it is \qty{559}{mW}.
    
    An even lower electrical power dissipation in CW operation is demonstrated from a ring QCSEL based on EV1299, with a diameter of \qty{100}{\um} and a waveguide width of \qty{3.1}{\um}, as visible from the LIV-characteristics shown in Fig.~\ref{fig:ring_CW}(a). The power dissipation is \qty{267}{mW} at threshold and is less than \qty{374}{mW} at maximum optical power. Notably, this device, with the epitaxial layer facing up, has been characterized on wafer level, as illustrated in the inset of Fig.~\ref{fig:ring_CW}(a). Further, the device is hosted by a 2-inch quarter wafer that is in vacuum contact with a temperature-stabilized heatsink.
    
    Finally, throughout the dynamic range of the laser, single-mode emission is observed and the resulting tuning of the peak wavenumber over the current is shown in Fig.~\ref{fig:ring_CW}(b). The mode can be tuned by \qty{-3}{cm^{-1}} and a tuning rate of \qty{-0.45}{cm^{-1}/mA} is deduced. The inset displays the spectrum acquired at \qty{28}{mA}, with a SMSR exceeding \qty{25}{dB}.


\section*{Conclusions}

    We have demonstrated QCSELs based on ring cavities supporting WGMs with footprints well below $(200\times200)$ \textmu m$^2$, operating at room-temperature.
    These footprints are comparable to those of VCSELs and enable the large-scale production of low-cost and low-power dissipating MIR lasers.
    To miniaturize ring cavities, it is essential to provide a low-loss boundary along which the WGM is propagating.
    We quantified the losses induced by this boundary in terms of an equivalent mirror reflectivity and then benchmarked the performance of the ring cavities against linear QCSELs.
    Superior reflectivities exceeding 0.95 have been found through experiments and lasing is expected from cavities with diameters below \qty{50}{\um}.
    Simulations suggest that reflectivities larger than 0.98 can be achieved employing an Al\textsubscript{2}O\textsubscript{3}/Au HR coating.
    Narrow width QCSELs have been demonstrated and for a \qty{100}{\um} diameter ring, the overall power dissipation in CW has been reduced to below \qty{374}{mW}.
    Future devices will incorporate a new contact arrangement to enable QCSEL arrays and larger outcoupling apertures with different grating designs to enhance the optical output power.
    Continuing to scale the cavity perimeter will ultimately guarantee single-mode selection and minimize the electrical power dissipation.

\section*{Acknowledgments}
    The authors gratefully acknowledge the financial support from Innosuise - Swiss Innovation Agency (Innovation Projects: 52899.1) and the ETH Zürich Foundation (Project: 2020-HS-348).
    The authors would like to thank Filippos Kapsalidis and Emilio Gini for their support during the device fabrication, as well as Mathieu Bertrand and Moritz Müller for further developing the automated wafer-level setup. Special thanks to Paolo Micheletti and Ina Heckelmann for their valuable inputs on the manuscript text.
    
\section*{Author declarations}
    
\subsection*{Conflict of Interest}
    The authors have no conflicts to disclose.

\subsection*{Intellectual Property}
    D.S. and J.F. have Patent EP4329113A1 pending.

\subsection*{Author Contributions}
    \begin{description}
        \item[D.S.] Conceptualization (supporting);
                    Data Curation (lead); 
                    Formal Analysis (lead):
                    Investigation (lead);
                    Visualization (lead);
                    Funding acquisition (supporting);
                    Writing - original draft (lead);
                    Writing - review and editing (equal);
        \item[M.B.] 
                    Resources (lead);
                    Writing - review and editing (equal);
        \item[J.F.] Conceptualization (lead);
                    Funding acquisition (lead);
                    Writing - original draft (supporting);
                    Writing - review and editing (equal);
                    Supervision (lead);
                    Project Administration (lead);

    \end{description}

\section*{Data availability}
    The data that support the ﬁndings of this study are available from the corresponding author upon reasonable request.


%
%

%

\clearpage
\bibliography{refs}

\end{document}